\documentclass[amsmath,amssymb,floatfix]{revtex4}
\usepackage{graphicx}
\usepackage{bm}
\begin{document}

\preprint{PRD}

\title[]{Comments on Accretion of Phantom Fields by Black Holes\\
and the Generalized Second Law}

\author{Jos\'e Antonio de Freitas Pacheco}
\affiliation{Laboratoire CASSIOP\'EE - UMR 6202 \\ Observatoire de la C\^ote d'Azur, \\ BP 4229
06304 Nice-France}
\email{pacheco@obs-nice.fr}
\date{\today}

\begin{abstract}

The thermodynamic properties of a phantom fluid and accretion by a black hole were recently
revisited by Pereira (2008) and Lima et al. (2008). In order to keep positive both the entropy and
the temperature, those authors assumed that the phantom fluid has a non null chemical potential.
In this short note we will show that there is a flaw in their derivation of the thermodynamic state 
functions which invalidates their analysis and their conclusions concerning the accretion of
a phantom fluid by a black hole.

\end{abstract}


\maketitle


In the past years a considerable number of papers have been addressed to the thermodynamic 
properties of phantom fields, particularly because they could be a possible driving agent  
of the observed accelerated expansion of the universe. A scalar field, supposed to be spatially
homogeneous, with a negative kinetic term is an example of a phantom field. This field is usually 
associated to a fluid of energy density $\rho$ and pressure $P$ by performing the
identifications $t_{00}=\rho$ and $t^a_a=3P-\rho$, where $t_{ab}$ is energy-momentum of the field. Under 
these conditions, the equation of state is $P=w\rho$, with the parameter $w<-1$ (generally assumed to 
be constant) being determined from the field potential $V(\phi)$. 

The thermodynamics of phantom fields are usually studied within the framework of a fluid model and
straightforward properties can be derived from the Euler's relation
\begin{equation}
Ts=(P+\rho)-n\mu=(1+w)\rho-n\mu
\label{euler}
\end{equation}
where $s$ is the entropy density, $T$ is the temperature, $n$ is the particle density and $\mu$ is the
chemical potential. Note that for ``classical" fields there is no relation defining the associated fluid particle 
density as a function of $\phi$ and $\dot\phi$ and, consequently, one assumes that the state functions depend 
only on the temperature, implying that $\mu=0$. In this case, since $w<-1$ it is trivial to show that
eq.~\ref{euler} requires either $s<0$ and $T>0$ or $s>0$ and $T<0$. If the second alternative is chosen in order
to keep the entropy positive, what is the meaning of a negative temperature? In fact, since 
$T=\left(\partial E/\partial S\right)$, the thermodynamic temperature can be interpreted as a measure of how the entropy 
varies as energy is injected or extracted from a given system.
In a phantom fluid, $s \propto \rho^{1/(1+w)}$, implying that if a cavity filled with such a fluid 
receives energy the entropy decreases. This singular behaviour is observed in experiments on the nuclear
spin relaxation of a LiF crystal after exposition in a magnetic field \cite{purcel51}.
Supposing that a phantom fluid satisfies the conditions $s>0$ and $T<0$ and the validity of the
Generalized Second Law, as proposed by \cite{beke74}, de Freitas Pacheco \& Horvath (\cite{pacheco07}) investigated 
the accretion of such a fluid by a black hole, concluding that there is a minimal mass above which the 
accretion process is not allowed.

Recently, Pereira (\cite{pere08}) and Lima et al. (\cite{lima08}) have revisited this problem, considering 
the possibility that the chemical potential of a phantom fluid be different from zero. In this case
(see eq.~\ref{euler}), if the chemical potential is negative, both the entropy and the temperature of the
phantom fluid can be positive quantities. In order to obtain the chemical potential, the aforementioned authors
assumed that the energy density and the pressure are functions not only of the temperature but also of the 
particle density. In this case, one can write
\begin{equation}
d\rho=\left(\frac{\partial\rho}{\partial n}\right)_Tdn + \left(\frac{\partial\rho}{\partial T}\right)_ndT
\label{dif1}
\end{equation}
and, combining this equation with the thermodynamic identity
\begin{equation}
\left(\frac{\partial\rho}{\partial n}\right)_T = \left[\frac{(P+\rho)}{n}-\frac{T}{n}\left(\frac{\partial P}{\partial T}\right)_n
\right]= \left[(1+w)\frac{\rho}{n}-\frac{wT}{n}\left(\frac{\partial\rho}{\partial T}\right)_n\right]
\label{parcial}
\end{equation}
one obtains
\begin{equation}
\dot\rho=\left[(1+w)\rho-wT\left(\frac{\partial\rho}{\partial T}\right)_n\right]\frac{\dot n}{n}+
\left(\frac{\partial\rho}{\partial T}\right)_n\dot T
\label{doteq}
\end{equation}
In the next step (see reference \cite{pere08}), the equations describing the evolution of the thermodynamic functions
in a Friedmann background were used, namely,
\begin{equation}
\dot\rho=-3(1+w)\rho\frac{\dot a}{a}~~~~~~ \dot n=-3n\frac{\dot a}{a}\,\,\,~~~~~~~  
\dot s=-3s\frac{\dot a}{a}
\label{cosmological}
\end{equation}
It should be emphasized that, as a consequence of the identification of the state functions with 
the trace and the ``time-time" component of the energy-momentum tensor of the scalar field, we have 
$\rho=\rho(\dot\phi,\phi)$ and $P=P(\dot\phi,\phi)$. However, there is no equivalent relation for the particle 
density. This problem was solved in reference \cite{pere08} by introducing ``ad hoc" the conservation law
equation $J^a_{;a}=nu^a_{;a}=0$. This, of course, is an acceptable procedure but conditioned to the abandon
of any association with the scalar field $\phi$. Replacing the cosmic evolution equations 
into eq.~\ref{doteq} one obtains easily
\begin{equation}
\frac{\dot T}{T}=-3w\frac{\dot a}{a}
\label{evol}
\end{equation}
Then, after some algebraic manipulations, the energy, entropy and particle densities are derived, i.e.,
\begin{equation}
\rho=\rho_0\left(\frac{T}{T_0}\right)^{(1+w)/w}~~~~~~ s=s_0\left(\frac{T}{T_0}\right)^{1/w}~~~~~~
n=n_0\left(\frac{T}{T_0}\right)^{1/w}
\label{resultados}
\end{equation}
where the subscript ``o" means reference values taken when the scale factor is equal to $a_0$. 
The chemical potential was finally derived from these relations and eq.~\ref{euler}, e.g.,
\begin{equation}
\mu=\frac{(P+\rho)}{n}-\frac{Ts}{n}=\left[(1+w)\frac{\rho_0}{n_0}-\frac{T_0s_0}{n_0}\right]\left(\frac{T}{T_0}\right)
\label{potencial}
\end{equation}

The first aspect to be remarked is that the resulting relations for the energy and the entropy densities
depend only on the temperature and not on the particle density and temperature, according to the
initial hypothesis made by those authors. This inconsistency is a consequence of the use of the cosmic
evolution equations (eqs.~\ref{cosmological}). These equations only say how the thermodynamic functions
vary as the universe expands and they are equivalent to an adiabatic expansion (in fact, the entropy
per unit comoving volume $sa^3$ is constant and if $\dot a = 0$ all thermodynamic functions remain constant). 
They should not be used to derive the thermodynamic functions since
the physical properties of the fluid do not depend on the state of expansion of the universe. 
In reality, the equation for the energy density derived in reference \cite{pere08} describe only
its variation during an adiabatic transformation, when the system goes from a state characterized by
a temperature $T_0$ to another of temperature $T$. In order to illustrate our statement, let us calculate
according to the recipe of \cite{pere08}, the properties of an ideal monoatomic gas. In this case,
the energy density and the pressure are functions of the particle density and temperature, as \cite{pere08,lima08}
have assumed and they are given respectively by $\rho = 3knT/2$ and $P=2\rho/3$ with $w=2/3$. Simple calculations
following the procedure by those authors give
\begin{equation}
\rho=\rho_0\left(\frac{T}{T_0}\right)^{5/2}\,\,\; n=n_0\left(\frac{T}{T_0}\right)^{3/2}\,\,\;
s=s_0\left(\frac{T}{T_0}\right)^{3/2}
\label{resul}
\end{equation}
As expected, the original energy density equation is not recovered and the derived relation
reproduces the well known result of how the energy density (or the particle density) varies during
an adiabatic transformation. The resulting expression for the entropy density is also uncorrect since
the entropy of an ideal gas is given by the Sakur-Tetrode formula, i.e.,
\begin{equation}
\frac{s}{k}=n lg\left[\left(\frac{e^{5/2}}{n}\right)\left(\frac{mkT}{2\pi\hbar^2}\right)^{3/2}\right]
\label{sakur}
\end{equation}  
Notice that since $T\propto a^{-2}$ and $n\propto a^{-3}$, eq.~\ref{sakur} satisfies 
the condition $sa^3=constant$. Moreover, the derived entropy density (eqs.~\ref{resul}), when compared to the Sakur-Tetrode 
formula, does not have the correct dependence either on the temperature or on the particle density. Similarly, had 
we adopted the procedure by
\cite{pere08,lima08}, the resulting chemical potential for the ideal gas would be
\begin{equation}
\mu=kT\left[\left(\frac{5}{2}-\frac{s_0}{kn_0}\right)\right]
\end{equation}
whereas the correct expression derived, for instance, from the free-energy $F$ is
\begin{equation}
\mu=kT lg\left[n\left(\frac{2\pi\hbar^2}{mkT}\right)^{3/2}\right]
\end{equation}

A possible solution of the thermodynamic equations for a fluid obeying an equation of state $P=w\rho$, having
a non null chemical potential is $\rho(n,T)=F(T)n^{(1+w)}$ and $s(n,T)=nG(T)$, where $F(T)$ and $G(T)$ are
arbitrary functions of the temperature. In this case, the chemical potential is given by
$\mu=(1+w)nF(T)-TG(T)$. As expected, for $w<-1$ the chemical potential is always negative and, as a consequence,
the entropy and the temperature are positive quantities. It is worth mentioning that using the cosmic evolution
equations, it results that the temperature remains constant for this fluid model, as the universe expands, 
while the energy density increases as $\rho\propto a^{-3(1+w)}$, a well known result.

Clearly, eqs.~\ref{resultados} and \ref{potencial} do not describe correctly the physical 
properties of a phantom fluid and, consequently, the analysis made in \cite{lima08} concerning the accretion 
of a phantom fluid by a black hole is inadequate and the conclusions by \cite{pacheco07} remain valid.


\begin{thebibliography}{}

\bibitem{purcel51}
E.M. Purcell and R.V. Pound, Phys.Rev. 81, 279 (1951)
\bibitem{pacheco07}
J.A. de Freitas Pacheco and J.E. Horvath, Class.Q.Grav. 24, 5427 (2007)
\bibitem{beke74}
J.D. Bekenstein, Phys.Rev.D 9, 3292 (1974)
\bibitem{pere08}
S.H. Pereira, astro-ph/0806.3701
\bibitem{lima08}
J.A.S. Lima, S.H. Pereira, J.E. Horvath and D.C. Guariento, astro-ph/0808.0860

\end{thebibliography}
\end{document}